\documentclass{iaus}
\usepackage{graphicx}

\title[Synoptic Sky Surveys] %% give here short title %%
{Exploring the Time Domain With Synoptic Sky Surveys}

\author[S.G. Djorgovski et al.]   %% give here short author list %%
{S.G.~Djorgovski$^{1}$, A.A.~Mahabal$^1$, A.J.~Drake$^1$, M.J.~Graham$^1$,
C.~Donalek$^1$, R.~Williams$^1$}
\affiliation{$^1$California Institute of Technology, Pasadena, CA, USA\\
email: {\tt \{\{george,aam,donalek\}@astro,\{ajd,mjg\}@cacr,roy@ligo\}.caltech.edu} \\[\affilskip]
}

\pubyear{2012}
\volume{285}  %% insert here IAU Symposium No.
\pagerange{1-3}
\jname{New Horizons in Time Domain Astronomy}
\editors{R.E.M.~Griffin, R.J.~Hanisch \& R.~Seaman, eds.}
\begin{document}

\maketitle

\begin{abstract}
Synoptic sky surveys are becoming the largest data generators in astronomy, and
they are opening a new research frontier, that touches essentially every field
of astronomy.  Opening of the time domain to a systematic exploration will
strengthen our understanding of a number of interesting known phenomena, and
may lead to the discoveries of as yet unknown ones.  We describe some lessons 
learned over the past decade, and offer some ideas that may guide strategic 
considerations in planning and execution of the future synoptic sky surveys.
\end{abstract}

% 12345678901234567890123456789012345678901234567890123456789012345678901234567890

\bigskip
\noindent{\large \bf 1.  Introduction: Exploring a New Domain}
\smallskip

In 1990's, astronomy transitioned from a data poverty to an immense, exponentially 
growing data richness.  The main agents of change were large digital sky surveys, 
that produced data sets measured from a few to a few tens of Terabytes, and they, 
in turn, were enabled by the information technology.  The challenge of the effective 
scientific use of such data sets was met by the advent of the Virtual Observatory 
(VO) concept.  The data volume continues doubling on a scale of $\sim 1 - 2$ years, 
reflecting the Moore's law that describes the growth of the technology that produces 
the data.  There is also an accompanying growth of data complexity, and data quality.  
We are now transitioning into the Petascale regime, and the main agents of change are 
synoptic sky surveys, that cover large areas of the sky repeatedly.  Some of the current 
surveys include CRTS, PTF, and PanSTARS in the optical, and various SKA prototypes in 
radio, leading to the next generation of facilities that will effectively operate in a 
time-domain mode, producing tens of TB daily, e.g., the LSST and SKA; and many others 
described at this conference.

The time domain astronomy (TDA) opens a new discovery space, not just by the sheer 
growth of data rates and data volumes, but also in opening the ``time axis'' (actually, 
many axes) of the observable parameter space (OPS).  A distinction should be made 
between the OPS, which is limited by our technology and the physical limitations
of measurements (e.g., transparency of the Earth's atmosphere or the ISM, diffraction
limit, quantum limits, etc.), and the physical parameter space, which is populated
according to the laws of nature; the mapping of one to the other is not trivial.
This expresses the vision of a 
systematic exploration of the OPS first formulated by Zwicky (1957), who referred to 
it as the ``morphological box''.  History has shown that every time that technology 
enables us to open a new portion of the OPS, we are likely to discover some new types 
of objects and phenomena (Harwit 1975).  Specifically, exploration of the time domain 
(``monitoring sky for variability'') was eloquently advocated by Paczynski (2000).

It is a rich territory to explore.  Some phenomena can be studied {\it only} in the 
time domain, e.g., various cosmic explosions, accretion and relativistic phenomena, 
etc.  As a whole, TDA touches essentially every field of astronomy, from the Solar 
System to cosmology, and from stellar structure and evolution to extreme relativistic 
phenomena.  Nor is it confined to the electromagnetic signals, as the neutrino and 
cosmic ray astronomy mature, and gravitational wave astronomy is born.  This very 
richness makes the TDA a too diffuse concept, just as it makes little sense to talk 
about the ``spectroscopic astronomy'' or ``imaging astronomy''.  Rather, we can more 
meaningfully focus on the subjects of synoptic sky surveys or transient event discovery
and characterization.

Recent discoveries of the previously predicted phenomena, e.g., supernova breakout
shocks, or tidal disruption events illustrate the scientific potential of TDA.  
It is reasonable to expect that a systematic exploration of the previously poorly 
covered parts of the OPS, in terms of the sensitivity, time cadences, area coverage, 
etc., may lead to a discovery of the previously unknown phenomena.

TDA was also recognized as one of the most promising areas of the new, data-rich 
astronomy at the very onset of the VO concept (Djorgovski et al.~2001ab), and indeed 
it exercises every envisioned VO functionality, and then some.  As we argue below, 
a strong computational infrastructure is an essential enabling factor for the TDA.

%%%%%%%%%%%%%%%%%%%%
\bigskip
\noindent{\large \bf 2.  Some Lessons Learned}
\smallskip

The field is far too big to review adequately here.  Our own
experiences may be illustrative of the challenges involved,
at least in the visible wavelength regime.

A search for highly variable and transient sources in the DPOSS plate overlaps 
Mahabal et al.~2001; Granett et al.~2003)
covered $\sim 8,000$ deg$^2$ with at least 6 exposures (2 in each of the 3 filters), 
and time baselines ranging from a few months to $\sim 8$ years.  We found that at 
these time baselines roughly a half of the high-amplitude variable objects are
Galactic stars (mainly CVs and flaring dM), and a half are AGN (mostly blazars).
We also found that in a single snapshot, there will be $\sim 10^3$ optical
transients per sky down to $\sim 20$ mag, and estimate that has held well since then.
It was clear that a variety of phenomena contribute to the population of
optical transients (OTs), but that (near) real-time follow-up observations
would be necessary to establish their nature.

The Palomar-Quest (PQ) survey (Djorgovski et al.~2008) and the concurrent NEAT project 
lasted $\sim 5$ years, ending in September 2008, 
with exploration of the time domain as the main science
driver.  It resulted in discoveries of several hundred supernovae (SNe), mostly
in collaboration with the LBNL Nearby Supernova Factory, studies of AGN
variability, and studies of the most variable sources on the sky (aside 
from SNe), that again turned out to be mainly CVs and beamed AGN (Bauer et al.~2009).  
For the last 2 years of the survey, we processed the drift scan observations in
real time, the PQ Event Factory, that in some cases led to a follow-up
spectroscopy within an hour of the initial OT detection.  The scientific returns 
were limited mainly by the poor quality of the data, and by the available follow-up.  
PQ was succeeded at the same telescope by the PTF (Rau et al.~2009), that operates with 
a very similar model, but with a much better camera and much more abundant 
follow-up resources.

Aside from the confirmation that an OT event stream will contain a broad
variety of astrophysical phenomena, several key lessons emerged.  
First, that asteroids are the main contaminant,
with $\sim 10^2 - 10^3$ asteroids for each astrophysical transient, and thus
a joint data processing and analysis is necessary.  
Second, that the adequate
follow-up -- and spectroscopy in particular -- is essential for the scientific
returns; this is still a critical issue, and it is getting worse.  
Third, that
rapid classification of transients is essential in order to distill the
incoming event stream down to a manageable number of interesting events worthy
of the expenditure of the limited follow-up resources.  A part of this is a
reliable and robust elimination of various data artifacts: in a massive data
stream, there will be inevitably many glitches, and even the most unlikely
things will happen, and most of them can look like transient events to a
data pipeline.  
And finally, that the cost of the software development will dominate 
any current or future synoptic sky surveys, accounting perhaps to $\sim 80$\% 
of the total cost.  One practical lesson was that the real-time processing
demands must be accommodated in the overall system architecture, in addition
to all that has been learned in the processing of single-epoch surveys.

We are currently conducting the Catalina Real-Time Transient Survey (CRTS;
Drake et al.~2008, Djorgovski et al.~2011a, Mahabal et al.~2011,
Drake et al., this volume).
CRTS taps into a data stream used to search for NEO astroids, thus both
satisfying the need to separate asteroids from astrophysical OTs, and 
illustrating yet again that the same data stream can feed many different
scientific projects.  CRTS has so far discovered 
$\sim 1,000$ SNe, including some novel or unusual types, 
a comparable number of CVs and dwarf novae, 
variability-based IDs of previously unidentified $Fermi$ $\gamma$-ray sources,
planets or other low-mass companions around white dwarfs,
young stellar objects, 
and a plethora of variable stars and AGN (see, e.g., Drake et al.~2010, 2011, 2012).
CRTS imposes a very high detection threshold for OTs, and even this subset 
of the highest amplitude events strains our follow-up capabilities.
If we modify the pipeline to pick all statistically significant 
variables, the number of OTs would grow by at least an order of magnitude.

We are accumulating an unprecedented data set of images and source catalogs
(light curves) for $> 5 \times 10^8$ sources covering $\sim 33,000$ deg$^2$,
spanning the time baselines from 10 min to $\sim 7$ years and growing.
This archival information is extremely useful for the interpretation of
OTs, and it can enable a variety of archival TDA studies.

One lesson of CRTS is that a synoptic sky survey need not be photometric:
its job is to discover transients, which can be done very efficiently
in a single bandpass (or just an unfiltered CCD); their photometry is 
best done as a part of the follow-up.  This relaxes many calibration
and data quality demands faced by the surveys that aim to be photometric.
One should separate discovery of OTs from their characterization.

Another, iterated lesson is that the spectroscopic follow-up is already
a key bottleneck, with only maybe $\sim 10$\% of CRTS transients followed.
This problem will get worse by orders of magnitude with the next
generation of synoptic sky surveys.  Thus, the need for an effective
automated classification of transient events is critical.

%%%%%%%%%%%%%%%%%%%%
\bigskip
\noindent{\large \bf 3.  Cyber-Infrastructure for Time Domain Astronomy}
\smallskip

TDA is by its nature very data-intensive, requiring a strong
cyber-infrastructure that includes data processing pipelines, archiving,
automated event classification and distribution, assembly of the relevant
information from the new data and the archives, etc.  

The ephemeral nature of transient events requires that they are electronically
distributed (published) in real time, in order to maximize the chances of a
necessary follow-up.  To this effect, we developed VOEvent, a VO-compliant 
standard for the event information exchange.
Our vision was to lay the foundations for the robotic telescope networks
with feedback, that would discover and follow-up transients, involving a 
variety of computational and archival data resources, and to facilitate 
event publishing, brokering, and interpretation.
The next step was to develop a concept of event portfolios, that would
automatically accumulate the relevant information and make it both
machine- and human-accessible, via the web services and various
electronic subscription mechanisms.  The current implementation is
$SkyAlert$ (Williams et al.~2009).

The challenge of an automated event classification and follow-up 
prioritization is still outstanding.  All OTs look the same when
discovered -- a star-like object that has changed its brightness significantly
relative to the comparison baseline -- and yet, they represent a vast
range of different physical phenomena, some of which are more interesting
than the others.  Nowadays, surveys generate tens to hundreds of OTs per
night; LSST may find $\sim 10^5 - 10^7$ per night.  Which ones are worthy 
of the expenditure of valuable and limited follow-up resources?

This entails some special challenges beyond traditional automated classification
methods, which are usually done in some feature vector space, with an 
abundance of homogeneous data.  
Here, the input information is generally sparse and heterogeneous, and often
with a poor $S/N$; there are only a few initial measurements, that differ 
from case to case, with differing measurement errors; the contextual information
is often essential, and yet difficult to capture and incorporate in the classification 
process; many sources of noise, instrumental glitches, etc., can masquerade 
as transient events in the data stream; new, heterogeneous data arrive, and 
the classification must be iterated dynamically.  The process must be
automated, robust, and reliable, with at most a minimal human intervention.
Requiring a high completeness (not missing any interesting events) and a low 
contamination (a few false alarms), and the need to complete the classification
process and make an optimal decision about expending valuable follow-up resources
in a (near) real time are substantial challenges that require some novel approaches
(Donalek et al.~2008, Mahabal et al.~2008, Djorgovski et al.~2011b). 

Most of the information about any given event initially, and often permanently,
would be archival and/or contextual: spatial (what is around the event), temporal
(what is its past light curve), and panchromatic (has it been detected on other 
wavelengths).  Applying it may require a human (expert) judgment, and yet,
human involvement does not scale to the forthcoming event data streams.
We are working on the methods to harvest the human pattern recognition
skills and turn them into computable algorithms.

%%%%%%%%%%%%%%%%%%%%
\bigskip
\noindent{\large \bf 4.  Concluding Comments}
\smallskip

TDA -- or simply astronomy with synoptic sky surveys -- is intrinsically an
{\it astronomy of telescope-computational systems}.  An 
optimal strategy may be to have dedicated survey telescopes, and surveys
that are not overburdened by other requirements, e.g., multicolor photometry,
and a hierarchy of follow-up facilities.  For example, there may
be a set of smaller, robotic telescopes providing a multicolor photometry
and helping select the most promising events for spectroscopy.
It would also make sense to coordinate surveys at different wavelengths
to serve as a first-order mutual multi-wavelength follow-up.

There is an understandable trend to optimize a given survey's parameters,
e.g., cadence, depth, etc., for a given scientific goal, e.g., SNe or NEO
asteroids.  That inevitably introduces selection biases against objects
whose variability may not be captured well with that particular window
function, thus diminishing the likelihood of truly novel discoveries.
It would be good to have a broad spectrum of time baselines that can 
capture a variety of phenomena, both known and as yet unknown.
It would make sense if the competing surveys would coordinate their 
sky coverage and cadences, and share the data.

An adequate and effective follow-up, especially the optical spectroscopy,
remains a key limiting factor.  In the regime where there is a steady and 
abundant stream of events, highly disruptive Target-of-Opportunity approach
is not optimal; dedicated follow-up facilities are needed.
The current generation of spectrographs 
at large telescopes tend to be optimized for a highly multiplexed 
spectroscopy of faint objects, e.g., for the studies of galaxy evolution.
In contrast, follow-up of transient events requires an efficient
single-object spectroscopy with relatively short exposures.  Trying to
repurpose the existing equipment for a highly inefficient use makes
little sense: telescopes and instruments dedicated for a spectroscopic 
follow-up of transient events should be designed accordingly.

All this has to be built on a strong cyber-infrastructure for data
processing and archiving, event discovery, classification and publishing,
etc.  Automated, robust, and reliable event classification is a key need
for effective scientific returns, and an optimal use of the expensive
facilities.  Given the importance of the archival data for the early
classification and interpretation of events, efficient, VO-type data
services will be increasingly more important.  Overall, a strong investment 
into astroinformatics, including facilities, software, and scientist training, 
is a major strategic need.

A transition from the data poverty regime to the data overabundance will also 
change the astronomical sociology and operational modes: we are already in the 
regime where the producers of these massive data streams cannot fully exploit 
them in a timely manner.  Thus, the focus of value shifts from the ownership 
of data to the ownership of expertise needed to make the discoveries.  A key 
concept, promoted by the CRTS, is the completely {\it open data} philosophy: 
making the synoptic sky survey and event data streams available immediately 
to the world.  While this trend was already apparent with the single-epoch 
surveys, it becomes critical with the synoptic sky surveys and the highly 
perishable transient events they discover.  As the data rates exceed the 
capabilities of any individual group to follow up effectively, it only makes 
sense to open them up, and thus engage a much broader segment of the astronomical 
community; in fact, it would be irresponsible to do otherwise.  While the concept 
of a proprietary data period may still make sense for some types of targeted 
observations, it does not for the exponentially growing data streams today 
or in the future.

Finally, perhaps the real-time astronomy with OTs is being overemphasized.
There is a lot of excellent, not time-critical science that can be done 
with the growing archives from synoptic sky surveys, e.g., a systematic
search for variables of given types (e.g., RR Lyrae for the mapping of
the Galactic structure), an improved characterization of AGN variability
as a constraint on theoretical models of accretion and beaming, etc.

\smallskip
\noindent{\bf Acknowledgments:}  
This work was supported in part by the NASA grant 08-AISR08-0085, 
and the NSF grants AST-0407448, AST-0909182 and IIS-1118041.
We thank numerous collaborators from the DPOSS, PQ, and CRTS surveys, and the
VO and astroinformatics communities for the work and ideas that have shaped
our ideas and scientific results.  We also thank the organizers for the
most interesting conference.

%%%%%%%%%%%%%%%%%%%%
\bigskip
\noindent{\large \bf Appendix: A Figure of Merit for Synoptic Sky Surveys}
\smallskip

It has become customary to compare surveys using the etendue, a product of the telescope
collecting area $A$ and the instrument field of view $\Omega$ as a figure of merit (FoM).  
However, $A \Omega$ simply characterizes the telescope and
partly the instrument, and says nothing about the survey, e.g., the depth, coverage rate,
cadence, etc.  A more meaningful FoM is needed.

We propose the following indicator of a survey's discovery potential, a product
of its spatio-temporal coverage rate, $C$, and the estimate of the depth, $D$, that
may be reasonably expressed as proportional to the $S/N$ ratio of the individual
exposures.  Thus:
$$ C = R \times N_p \times f_{open} ~~~{\rm and}~~~
   D = \left[ A \times t_{exp} \times \epsilon \right] ^{1/2} / FWHM ~\sim S/N $$
\noindent
where 
$R$ is the area coverage in deg$^2$/night (not counting repeated exposures), 
$N_p$ is the number of passes per field in a given night, 
$f_{open}$ is the fraction of the open time averaged over the year, including the
 weather losses, engineering time, deliberate closures, etc.,
$A$ is the effective collecting area in $m^2$,
$t_{exp}$ is the average exposure time in sec,
$\epsilon$ is the overall efficiency (throughput) of the instrument, and
$FWHM$ is the typical seeing FWHM in arcsec.
Clearly, all these parameters should be taken as typical or averaged over a year.
Note that $f_{open}$ and $FWHM$ characterize the site,
$A$ and $\epsilon$, and partly $R$, characterize the telescope+instrument,
and the remaining parameters reflect the chosen survey strategy.

$CD$ represents a FoM for a discovery rate of events, and net discovery potential
of a given survey would be $CD$ multiplied by the number of years the survey operates.
While this FoM accounts for most of the important survey parameters, it still does
not capture the factors such as the sky background and transparency, the total number
of sources detected (which clearly depends strongly on the Galactic latitude), the
cadence, the bandpasses, the angular resolution, etc.; nor it accounts for the
operational parameters such as the data availability, the time delay between the
observations and the event publishing, etc.  Nevertheless, we believe that $CD$
is a much more relevant FoM than the traditional (and often mis-used) $A \Omega$,
as far as a characterization of {\it surveys} is concerned.

The following table shows the estimated values of the relevant parameters and $CD$
for the 3 component of the CRTS, and several other current or future surveys.  
The assumed values of input parameters are based on our own experience or on the 
published values, and may be consistently too optimistic.  
The values of $CD$ are no better than $\sim 20$\%.

\smallskip
\begin{center}
\begin{tabular}{|l|c|c|c|c|c|c|c|c|c|c|}\hline
{\bf Survey} & {\bf $R$} & {\bf $N_p$} & {\bf $f_{open}$} & {\bf $A$} & {\bf $t_{exp}$} & {\bf $\epsilon$} & {\bf $FWHM$} & {\bf C} & {\bf D} & {\bf CD} \\ \hline
% & {$deg^{2}$} &  &  & {$m^{2}$} & {sec} & & {''} &  &  & \\ \hline
CRTS:CSS & 1200 & 4 & 0.7 & 0.363 & 30 & 0.7 & 3 & 3360 & 0.92 & 3090\\
CRTS:MLS & 200 & 4 & 0.7 & 1.767 & 30 & 0.7 & 3 & 560 & 2.03 & 1140\\
CRTS:SSS & 800 & 4 & 0.7 & 0.196 & 20 & 0.7 & 3 & 2240 & 0.55 & 1240\\
CRTS total & 2200 & 4 & 0.7 & (2.326) &  & 0.7 & 3 & 6160 &  &5470\\
PTF & 1000 & 2 & 0.7 & 1.131 & 60 & 0.7 & 2 & 1400 & 3.45 & 4820\\
SkyMapper & 800 & 2 & 0.7 & 0.785 & 60 & 0.8 & 2 & 1120 & 3.07 & 3440\\
PS1 & 1000 & 4 & 0.7 & 2.54 & 30 & 0.8 & 1 & 2800 & 7.81 & 21860\\
LSST & 5000 & 2 & 0.75 & 34.9 & 15 & 0.8 & 0.8 & 7500 & 25.6 & 192000\\
\hline
\end{tabular}
\end{center}

\end{document}